# GRAHAM'S FORMULA FOR VALUING GROWTH STOCKS

By Dr Andreas A. Aigner & Walter Schrabmair

*Benjamin Graham introduced a very simple formula for valuing a growth stock in 1962. How does it work and why? What is a sensible way to calculate this across many stocks and provide a scoring system to compare stocks amongst each other? We are presenting a methodology here which is put into practice.*

Benjamin Graham [1] is indisputably regarded as the father of value investing. He had quite an impressive life and was the mentor of many other investment greats. Born as Benjamin Grossbaum in London 1894 his family moved to the US at age 1. The market turmoil during WWI put his family through hardship but Benjamin was a gifted student and finished school with offers to start teaching.

He declined because he wanted to work on Wall Street, that's why he also changed list last name from Grossbaum to Graham, since it was kind of a disadvantage to have a German last name right after WWI. He started working as a paper boy and was soon promoted due to his skills at work and his numerical aptitude and not least due to his inherent interest in valuing companies on his own. At age 29 he left to start his own firm and at age 34 taught at Columbia University.

His most famous works include the 1934 book 'Security Analysis' [2] followed by 'The Intelligent Investor' in 1949 [3, 4]. His first book is a structured analysis of all products at the time while his second book 'The Intelligent Investor' aims more at the psychological & behavioral aspects of the stock market. He was a great teacher, so his books reflect this and are written very well and translated world-wide. While his insight into valuations of companies and the stock market is his own, there was already a lot of ongoing research and publications in this field before his seminal book 'Security Analysis' as Dennis Butler (FinancialHistory.org [2006]) [5] argues (See the references below that we have provided pre-dating Graham's work [6-27])

> A guide for the general reader for wise investment practice under today's financial conditions. (Security Analysis, Book Cover 1934)

While his work is based around a number of techniques to value companies using a number of metrics, the majority of his work centers around companies with a decent history of a balance sheet and valuation metrics in order to make sound investment choices which provide the biggest return while minimizing capital at risk. Yet there is one chapter from his 1962 edition of "Security Analysis" entitled

> "Newer Methods for Valuing Growth Stocks"

that leads us into his view on growth stocks [28]. This chapter 39 was removed subsequently from later editions [29, 30].

# WHAT IS THE DEFINITION OF A GROWTH STOCK?

> A growth stock is a share in a company that is anticipated to grow at a rate significantly above the average for the market. These stocks generally do not pay dividends, as the companies usually want to reinvest any earnings in order to accelerate growth in the short term. Investors then earn money through capital gains when they eventually sell their shares… (Investopedia 2019) [28]

In other words, as an investor you are betting purely on the success and future growth of such a company. You risk your capital without hoping for a payout in earnings in the future and simply hope to sell your shares for a profit, that is, if you do not hold the shares forever or until a stage much later in your life (e.g. to retire, to buy a house, a car etc.).

Why would you prefer to invest in a growth company instead of a company that pays you a steady income through dividends, by way of its earnings income which is very predictable and well known for many years? A Telco company is a prime example. Future growth is minimal, maybe 2% for example, but their quarterly/yearly earnings are massive and they can afford to pay out more than 75% of earnings as dividends. Take AT&T which pays out 70% of earnings yielding around 6.85% annually currently. What makes you choose a growth stock over a stock like this? It makes sense to compare returns to the most secure asset around: treasuries. Its yield is commonly known as the *risk-free rate*. Let's take the 30-year risk-free rate 3%. You can invest your cash returning 3% annually or you can buy shares in a stable company returning dividends, let's say 4–10% annually, or you invest your money in companies growing in excess of and sufficiently beyond your stable dividend yields. This margin is what Graham referred to as the *margin-of-safety.* It is the prime factor in determining whether it is more suitable to invest in any of the three categories at any time. What are growth rates offering a sufficient *margin-of-safety*? Let's take a simple example and look at this table of returns and their annually compounded return, Figure 1.

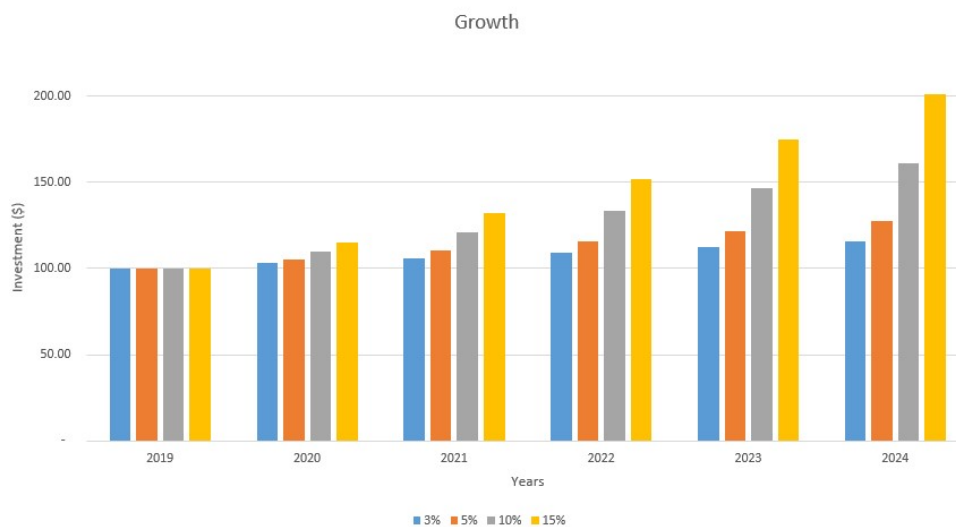

**Figure 1: Examples of compounded growth**

Starting with 100$ today with an annual return of 15% you have 201.14$ after 5 years while you have 127.62$ with a 5% return and only 115.93$ with a 3% return.

If you are invested in such a growth stock yielding 15% annually great! But of course the risk of the growth stopping, stagnating or just being plain wrong is there. How can you aim to minimize this risk or find an optimal way to maximize returns?

# BENJAMIN GRAHAM'S FORMULA (CHAPTER 39, SECURITY ANALYSIS, 1962)

Benjamin Graham gives us a formula to forecast the future value of a company based on its earnings history and growth forecasts. He compares it to other methodologies at the time, the Molodovsky method (1958) [31] and other more complicated methods. He refers to his method being based on an approach independently developed by Charles Tatham in a company publication of Bache & Co on 2 Oct 1961 and chapter 43 of a book by the same author (which is unfortunately missing a reference).

Benjamin Graham's formula [32] given in his book 'Security Analysis' (1962 version, Ch 39) is quite a simple formula to forecast the future value of a stock and is given by

$$(8.5 + 2 \cdot G) \cdot E = \frac{P}{E} \cdot E = Intrinsic\ Value \qquad \text{(Eq 1)}$$

G is the 'Growth in percentage points' and E are the 'Earnings' in local currency. We have abbreviated the factor by P/E since it is really a PE (Price/Earnings Ratio). The product of P/E and Earnings equates to the price of a company. The numbers in the P/E factor according to Graham derive simply from the assumption that a fair P/E of a company of <u>zero growth</u> is 8.5 and a company with a 2% growth should have a P/E of 12.5.

If you insert the forecast growth and earnings for the next year you get a price forecast for that year ahead. If you insert the current year growth and earnings forecast you get the current year price forecast, which is what Graham calls *'intrinsic value'*. We can calculate the 5 year forecast using a growth forecast for 5 years and earnings forecasts for 5 years from now, Figure 2.

The growth forecast can be modeled off the market, the industry of the stock and subjective interpretations of the future and the stocks growth. We like to use the average of forecasts that <u>analysts of the stock</u> have come up with. They all use different methods and ways to come up with a growth forecast of the stock which they will use to value the stock in their analysis. We want to take the average of all these analysts since it will filter out any kind of erroneous model and arrive at a consensus growth for the stock without us having to make a detailed model of the economy or industry.

Graham suggests to make an empirical forecast of earnings several years ahead but normalizing earnings by adjusting for market booms and/or recessions and/or high / low earnings. Here we model the earnings off the past earnings history, current year and next year earnings forecasts. We simply <u>linearly interpolate</u> 5 years ahead using a least-squares method. There are several ways one can do this and improve this method, using a longer time history or any other more sophisticated time series modelling technique [33]. For now, we just linearly interpolate the earnings to 5 years out.

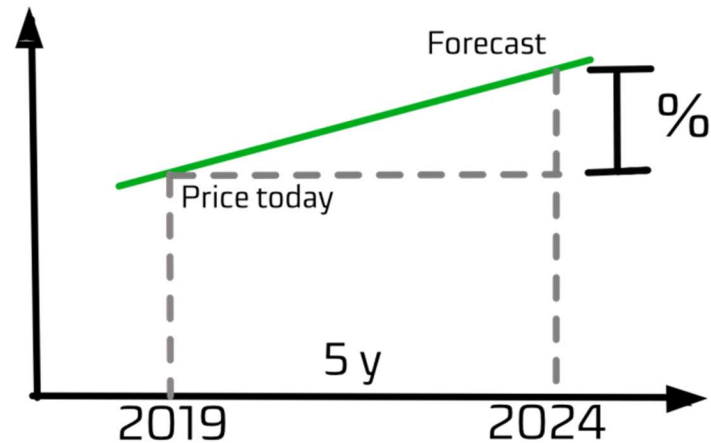

Figure 2 : Extrapolating the stock price

That Benjamin Graham found zero growth companies to have a P/E floor of 8.5 is not so farfetched. If we look at companies with near zero growth, we will see a P/E around 10. It won't be 8.5 necessarily but it is quite close to Benjamin's level. It is most likely related to the size of earnings and the dividend it pays out, in the end the price you pay for a stock in terms of earnings means that after 10 years the company will have earned as much as its current stock price is worth. If it pays out all these earnings as dividends it means after 10 years you have all your money back (pre-tax). The P/E won't go much lower because 1) a P/E of 0 would mean the stock price is zero or 2) a P/E of 1 would mean the stock is worth as much as its earnings. Both rather unlikely normally. The P/E of 8.5 represents the opportunity of earning above risk-free rates plus extra for any possible growth in the future.

## IMPROVEMENTS & LIMITATIONS TO THE MODEL

It's obvious to anyone that a model like this is limited and maybe even naive. No doubt one could build even more complicated models to forecast and find current intrinsic value. One limitation of such a model is related to the *'Petersburg Paradox'* [27], which is the problem that the value of a company with endless growth and earnings should be infinite. There <u>has</u> to therefore be a sensible approach to valuing companies beyond any significant time-frame.

There are also a number of ways it can be improved, Graham ignores discounting and interest rates on purpose. Furthermore, deriving a forecast from a set of analysts gives us more statistical metrics to include in our forecast and elaborate on the accuracy and significance of the forecast. The list of improvements is numerous. However, we like to keep it simple to differentiate between thousands of stocks so a formula like Graham's is just perfect for number crunching.

## LET'S LOOK AT RESULTS!

Being equipped with a formula for forecasting growth how do we search through the masses of stocks available for investment?

First of all, Market Capitalization (Market Cap) is a good way to subdivide the global stock market into groups. After all it doesn't make sense to compare a 1 tn $ company to a 40 bn $ company. Commonly one differentiates between 6 different types (Source: Investopedia).

| | | | | | |
|---|---|---|---|---|---|
| Mega Cap  | $ 300 bn | ≤ | Market Cap | | |
| Big Cap   | $ 10 bn  | ≤ | Market Cap | < | $ 300 bn |
| Mid Cap   | $ 2 bn   | ≤ | Market Cap | < | $ 10 bn  |
| Small Cap | $ 300 m  | ≤ | Market Cap | < | $ 2 bn   |
| Micro Cap | $ 50 m   | ≤ | Market Cap | < | $ 300 m  |
| Nano Cap  |          |   | Market Cap | < | $ 50 m   |

Secondly we use the analyst forecasts for the growth estimate, so we want to set a minimum number of analysts for each of the 6 Market Caps

| | |
|---|---|
| Mega Cap  | 25 |
| Big Cap   | 20 |
| Mid Cap   | 15 |
| Small Cap | 10 |
| Micro Cap | 5  |
| Nano Cap  | 3  |

The choice is relatively arbitrary in that we know that Mega Cap stocks will have a big number of analysts covering the name. The number of analysts slowly decreases going from Big Cap to Small Caps, below which not many analysts will cover Micro or Nano Cap stocks. Who knows, those few analysts might just be looking at the next Big Cap or Mega Cap stock of the future, so we want to keep a minimum of 3 at least for Nano Caps, since we hope that the amount of 'madness' or 'mistakes' between the analysts is averaged out between the analysts.

We want to further limit our search to those stocks with a minimum of 15% annual 5-year growth estimate in addition to a growth greater than zero for the last 5 years. We sort the results by 5Y% growth in descending order.

Doing this we arrive for example at the top 10 Mega Cap stocks Figure 3.

**(WK 11) Graham's Growth MegaCap Buy/Sell**

| ▲BUY | 5Y% | 1Y% | 0Y% | P5% | AN# | CR | ▼SELL | 5Y% | 1Y% | 0Y% | P5% | AN# | CR |
|---|---|---|---|---|---|---|---|---|---|---|---|---|---|
| BABA | 121 | 15x | 419x | 30 | 37 | 1.8 | | | | | | | |
| FB | 35 | 2x | 4x | 42 | 51 | 4.4 | | | | | | | |
| JPM | 27 | 2x | -7 | 15 | 26 | 0 | | | | | | | |
| AMZN | 23 | 98 | -6 | 101 | 49 | 1.1 | | | | | | | |
| AAPL | 22 | 124 | 72 | 8 | 40 | 1.6 | | | | | | | |
| MSFT | 21 | 39 | 93 | 18 | 32 | 2.8 | | | | | | | |
| V | 21 | 92 | 28 | 22 | 37 | 1.3 | | | | | | | |
| GOOGL | 8 | 130 | 31 | 15 | 46 | 3.4 | | | | | | | |
| WMT | 5 | 12 | -22 | 1 | 28 | 0.8 | | | | | | | |

5Y%: 5-Year Ann Ret Est, 1Y%: 1-Year Ret Est, 0Y%: Curr Year Ret Est, P5%: Past 5Y Ret, AN#: Curr Year # of Analysts, CR: Current Ratio

Copyright TradeFlags 2019    These are predictions based on Earnings & Growth and intended for information purposes only    Sun Mar 15 18:53:17 CET 2020

**Figure 3 : Top10 Mega Cap stocks with a 5-year growth >15% and those <0%**

You notice we have a list of buys on the left and a list of sells on the right. We haven't mentioned sells so far. For the sells we want to filter for stocks where the 5-year growth forecast is negative and sort by ascending order.

You see from this list number 1 is Alibaba (BABA) with predicted whopping 121% annual 5Y growth, it has a past growth of 30% and 37 analysts covering the name. This seems crazy at this point so maybe the analysts are a little too bullish on this chinese Amazon. But you have to remember that with this current brutal selloff due to the corona virus the stock is down 22% from the peak which contributes a little bit to this number. I expect the analyst growth forecasts are overly bullish on this name, so one should take a very large margin-of-safety in that case.

Amazon (AMZN) with a predicted 23% annual 5Y growth, it has a past growth of 101% and 49 analysts covering the name. If you consider past growth, then 23% annual growth the next 5 years seems very low. Are the expectations due to the virus really that bad? How does the past evaluate given we now have the growth of the past 5 years? 5 years ago AMZN was trading at 292.24 so the 5 year PE = (8.5+2*101) = 210.5 multiply this with 2019 earnings which is 27.12 then equates to a Graham Estimate of 5708.76 for 2019. AMZN is trading around 1900 at the time of writing. So you see an estimate even given exact figures in the past is not exact, yet no one would have priced in an annual growth of 101%. If you take half of this growth (50%) you arrive at a target price of 2969, which is quite similar to analyst 1Y target prices today.

Compare this to Apple (AAPL) which has 22% predicted annual 5Y growth and a past growth of 8% and 40 analysts covering. This forecast seems completely reasonable and one has a 7% margin of safety to our hurdle rate of 15%.

Microsoft comes in with 21% predicted annual 5Y growth and a past growth of 18% and 32 analysts covering. Also completely in line with the past.

This search for Mega Caps gives you the top names that everyone mentions when they talk about growth stocks (in addition to some others of lesser fame).

To further differentiate between these names we can look at the CR column. The CR (CurrentRatio) is a liquidity ratio measuring the ratio of assets to liabilities, e.g. debt. Amazon, AAPL & V have a ratio of around 1 and GOOGL & MSFT have a ratio of around 3 and FB a ratio of 4. Meaning FB is roughly 4 times more liquid than Amazon and MSFT is around 2 times more liquid than AAPL.

How would a bar chart like our example growth bar chart above look like for these names 5 years going forward? Take a look at Figure 4.

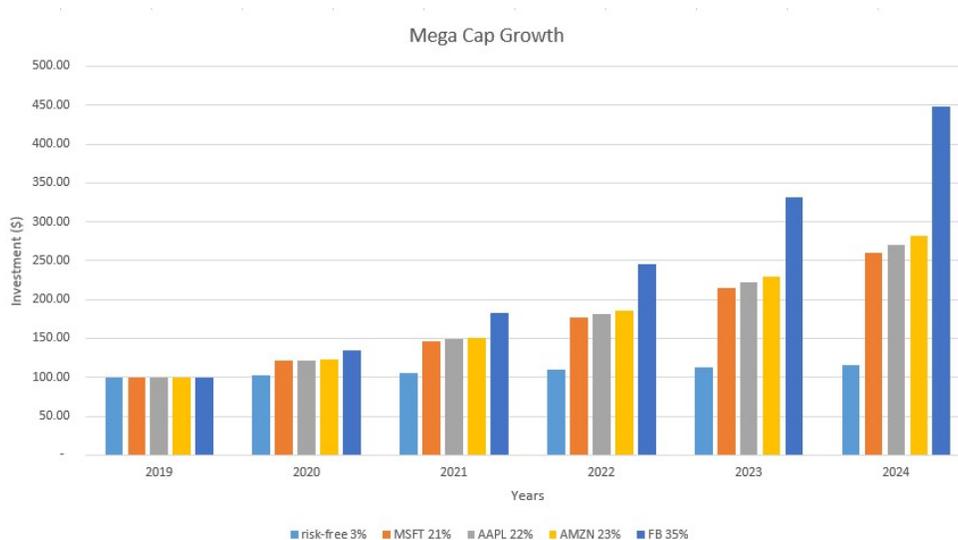

Figure 4 : Example Mega Cap Growth Chart for MSFT, AAPL, FB and AMZN

# CONCLUSION

We have provided you with a tool to give you a screen once a week of the top names that have the most growth or are candidates for sells. If you are interested in looking at the screens for the 5 other Market Caps please visit our Twitter, StockTwits, Reddit, Facebook Page or Homepage where we publish these details weekly.

We are providing a summary of growth figures across all names and major sectors & industries. Here is a sample screen showing the summary for the major sectors (left) and industries (right), together with an overall across all names (bottom right). There are three columns W5Y% the forecast for the next 5 years' annual growth, WP5% the past 5-year annual growth and #'s which is the number of stocks we are averaging over. How do we average? It doesn't make sense to average across all names equally to get a representative figure. These averages are <u>weighted according to their current market capitalization</u>; i.e. smaller possibly more aggressive growth companies have less weight than big possibly more conservative growth companies.

Since bigger market cap stocks tend to have more analysts covering it, the growth figures are more likely to be more 'sensible'. So it works both ways, in terms of weighting according to market cap and indirectly according to analysts. We also only include names with a minimum of 10 analysts. This removes names which are likely to have outdated analyst growth targets or which are too small to be representative. These figures therefore are 'conservative' averages and individual names which outperform these numbers should be attractive candidates. Companies with growth figures less than this average are probably unattractive candidates, unless one expects a strong revision to earnings or growth expectations, for example Biotech companies which might have just discovered something big or so-called disruptive companies in general.

The quickest way to find all the most recent Growth Summaries is to search for GSummary in the search field of our website, e.g. https://www.tradeflags.de/?s=gsummary.

### (WK 10) Graham's Growth Summary

| Sector | W5Y% | WP5% | #'s | Industry | W5Y% | WP5% | #'s |
|---|---|---|---|---|---|---|---|
| Technology | 20 | 17 | 147 | Software | 12 | 35 | 42 |
| Consumer Cyclical | 38 | 43 | 121 | Biotechnology | 21 | 24 | 31 |
| Industrials | 19 | 13 | 75 | Oil & Gas E&P | 13 | 21 | 24 |
| Healthcare | 21 | 13 | 97 | Specialty Retail | 52 | 64 | 22 |
| Financial Services | 22 | 15 | 83 | Diversified Industrials | 7 | 10 | 17 |
| Energy | 18 | 6 | 54 | Internet Content & Information | 63 | 26 | 18 |
| Basic Materials | 19 | 12 | 41 | Software Infrastructure | 18 | 16 | 19 |
| Consumer Defensive | 7 | 7 | 40 | Banks Regional US | 21 | 15 | 16 |
| Utilities | 3 | 5 | 21 | Semiconductors | 24 | 26 | 22 |
| Communication Services | 22 | 12 | 20 | All Sectors | 21 | 18 | 694 |

W5Y%: Weighted 5-Year Ann Ret Est, WP5%: Weighted Past 5Y Ret, #'s: Number of Stocks, we only select >=10 analysts
Copyright © TradeFlags 2019   These are predictions based on Earnings & Growth and intended for information purposes only   Thu Mar 12 16:40:43 CET 2020

**Figure 5 : Screen of Graham's Growth Summary across major sectors (left) and major industries (right). Showing market-cap weighted averages of 5Y annual growth in the future and past performance, together with the number of stocks in the population #'s**

**Biography:** *Dr Andreas A. Aigner has a PhD in Mathematics from Monash University, Melbourne, Australia, where he was born and studied. He spent a number of years in Research for various UK universities and worked almost 10 years for Morgan Stanley in Controlling, Trading & Pricing for the Exotic Derivatives desk in Hong Kong. He is now engaged full-time in research and is building a signaling automaton (TradeFlags) together with his longtime friend and associate Walter Schrabmair, who works at the Medical University of Graz and the Technical University of Graz in various research roles and as a computer whiz. Their contact emails are* andreas@tradeflags.at *and* walter@tradeflags.at